\documentclass[12pt]{iopart} 
\usepackage{epsfig} 
\usepackage{graphicx} 
\usepackage[english]{babel} 
\usepackage{setspace}

\begin{document}

\title{Approximate Model  of Neutron Resonant Scattering in a Crystal}

\author{Arnaud Courcelle$^1$, John Rowlands$^2$}

\address{$^1$ Commissariat \`a l'Energie Atomique, CEA/Cadarache, Bat 151, 13108 Saint Paul Lez Durance, France}
\address{$^2$ 81 South Court Avenue, Dorchester, Dorset DTI 2DA, United Kingdom} 
\ead{arnaud.courcelle@cea.fr, rowlandsjl@aol.com}

\begin{abstract}

In the theory of resonant scattering, the double differential cross section 
involves the computation of a multifold integral of a 4-point correlation function, 
which generalizes the traditional 2-point correlation function of Van-Hove 
for potential scattering. In the case of a neutron-crystal interaction, the numerical computation 
of these multifold integrals is cumbersome. In this paper, a new approximation is suggested. It is based on a factorization of the differential cross section into one function describing the exchange of kinetic energy between the neutron and the bound nucleus (phonons dynamic) and a function related 
to the nuclear scattering amplitude. This formalism is then applied to the modeling of 
resonant scattering of a neutron by $^{238}U$ in a $UO_2$ crystal lattice. 
\end{abstract}

\maketitle

\clearpage

\section{Introduction} 

The computation of the double differential scattering cross section (DDCS) 
at low incident neutron energies is required to solve neutron transport problems. 
Wigner and Wilkins \cite{wigner1944} used a two-body kinematic approach 
to study potential scattering in a free gas. Under the same classic assumptions, 
Blackshaw and Murray \cite{blackshaw1967} studied
the case of an energy-dependent cross section and further generalizations 
were investigated by Ouisloumen and Sanchez \cite{ouisloumen1991} and Rothenstein and Dagan \cite{roth4}. 

A general quantum formalism due primarily to Van Hove \cite{vanhove1954}
expresses the DDCS of potential scattering by a bound nucleus as a Fourier transform of a 
2-point correlation function.  Kazarnovski et al. \cite{kazarnovski1962}
 and Word and Trammell \cite{word1981}  extended the Van Hove theory to study resonant processes. 
The resonant DDCS becomes a Fourier double-Laplace 
transform of a 4-point correlation function.

In the case of a harmonic crystal, the numerical computation of the multifold integral is difficult 
because of the highly oscillating behavior of the 4-point correlation function.
A simplified multiphonon model, known as the "uncoupled phonon approximation" (UPA) 
was proposed by Naberejnev \cite{nab2001}. However, the validity of this approximation is questioned especially
within the short time range (case of high temperature for instance).
The present work suggests a new approximation that gives a correct limit at short time and provides a simple formula to compute 
the scattering kernel. 

\section{General formalism}

\subsection{The 4-point correlation function}

When the scattering amplitude is independent of incident neutron energy, it is known since the work of Van Hove \cite{vanhove1954} that the scattering DDCS 
is related to the Fourier transform of a two point correlation function $\chi_2(\tau)$ :

\begin{equation}
\frac{d^2\sigma}{d\Omega dE_f} = \frac{\sigma_p}{4\pi} \frac{k_f}{k_i} \mathcal{F}\{\chi_2(\tau)\}(\Delta E).
\end{equation}

$\Delta E = E_f - E_i$ is the energy transfer with $E_i$ and $E_f$ the energy of the neutron before and after the scattering event.
The Fourier transform is defined as
\begin{equation}
\mathcal{F}\{\chi_2(\tau)\} (x) = \frac{1}{2\pi} \int_{-\infty}^{\infty} e^{-i x \tau }\chi_2(\tau) d\tau.
\end{equation}

$\mathcal{F}\{\chi_2(\tau)\}(\Delta E )$ is the usual scattering function $S(\alpha,\beta) = S(\vec{ \Delta k },\Delta E)$ which depends on the momentum transfer $\vec \Delta k = \vec k_f - \vec k_i$ where $\vec k_i$ and $\vec k_f$ are the initial and final neutron wave vector respectively .
The theory was later generalized to treat the case of resonant scattering \cite{word1981}. In the incoherent approximation, the scattering cross section includes the resonant, the potential-resonant interference and potential terms. It can be demonstrated that
the DDCS of the resonant term for instance is related to the Fourier-double-Laplace transform $\mathcal{L}_2 \mathcal{F}$ 
of a 4-point correlation function $\chi_4(\tau,t,t')$:

\begin{equation}
\label{eq4p}
\fl \frac{d^2\sigma}{d\Omega dE_f} = \frac{\sigma_m}{4\pi} \frac{\Gamma_n}{\Gamma} \frac{k_f}{k_i}
 \frac{\Gamma^2}{4} \mathcal{L}_2 \mathcal{F} \{ \chi_4(\tau,t,t') \}(\Delta E, -i(E_i-E_0) + \Gamma /2, +i(E_i-E_0) + \Gamma /2).
\end{equation}

$E_0$, $\Gamma_n$ and $\Gamma$
are the energy, the neutron and total width of the resonance.
$\sigma_m = 4 \pi g \Gamma_n / k_i^2 \Gamma$
and $g$ is the usual statistical spin factor. The Fourier-double-Laplace transform is defined as 

\begin{equation}
\fl \mathcal{L}_2 \mathcal{F}\{ f(\tau,t,t') \} (x,z1,z2) = \frac{1}{2\pi} 
 \int_{-\infty}^{\infty}\int_{0}^{\infty}\int_{0}^{\infty} e^{-i x \tau } e^{ - z_1 t  } e^{- z_2 t'}\ f(\tau,t,t') d\tau dt dt'.
\end{equation}

Similarly, the resonant-potential interference term is expressed as a Fourier-single-Laplace transform of a 3 point correlation function $\chi_3(\tau,t)$. 

A quantum calculation shows that $\chi_4(\tau,t,t')$ is a function of time-dependant displacement operators:
 
\begin{eqnarray}
\chi_4(\tau,t,t') & = &  < \exp \left( -i\vec k_i. \vec r(\tau+t-t') \right) \exp \left( i\vec k_f.\vec r(\tau+t)\right) \nonumber \\
    & \times &  \exp \left( -i\vec k_f.\vec r(t)\right)  \exp \left( i\vec k_i.\vec r(0) \right) >_T.
\label{chi4}
\end{eqnarray}

Note that $\chi_2(\tau) = \chi_4(\tau,0,0)$ and $\chi_3(\tau,t) = \chi_4(\tau,t,0)$.
$<X>_T$ denotes the thermal average at temperature $T$. In this paper, time $\tau$, $t$ and $t'$ are expressed in unit of energy and $\hbar =1$.
The right hand side of Equation 1 and 3 should be multiplied by $\left( \frac{M+m}{m} \right)^2$ which is implicitly assumed in the paper.
$M$  and $m$ are the mass of the target nucleus and incident neutron and $A = M / m$ respectively . The scattering is assumed isotropic in the center-of-mass frame and  
multiple scattering effects are neglected (single collision approximation). Equation 3 assumes that the
Hamiltonian of the target and compound nucleus is the same. This condition neglects
the mass change of the target nucleus when the neutron is absorbed and re-emitted: $M >> m$. 

\subsection{Free-gas versus harmonic crystal-lattice model}

In the case where the target nucleus behaves like a free gas, the displacement operator is proportional to the momentum operator $r(t) = r(0) + p(0) t/M$. After averaging over a Maxwellian distribution of momentum at temperature $T$, $\chi_4$ within the free gas model (FGM) becomes

\begin{equation}
\label{eqfgm}
\fl \chi_4(\tau,t,t') = \exp\{ \frac{-i }{2 M}[ \tau (\vec k_f - \vec k_i)^2 + ( t - t' ) k_i^2] 
- \frac{kT}{2 M} [-\vec k_i ( t - t') + \tau ( \vec k_f - \vec k_i ) ]^2 \}.
\end{equation}

When the target nucleus is bound to a crystal lattice, the 4-point correlation function is more complex.
Under the harmonic approximation, the Bloch theorem \cite{nab2001} transforms $\chi_4$ into

\begin{eqnarray}
\label{eqclm}
\chi_4(\tau,t,t')  = \exp\{ -\rho ( \vec k_i,\vec k_i,0)  - \rho ( \vec k_f,\vec k_f,0 ) \} \nonumber \\
                     \times \exp\{ \rho ( \vec k_i,\vec k_f,-t' ) + 
		  \rho ( \vec k_f,\vec k_i,+t ) + \rho ( \vec k_f,\vec k_f,\tau)  \}  \nonumber \\
                      \times \exp\{ \rho ( \vec k_i,\vec k_i,\tau-t'+t) - 
		  \rho ( \vec k_i,\vec k_f,\tau-t' ) -  \rho ( \vec k_f,\vec k_i,\tau+t ) \}. 
\end{eqnarray}

For a cubic lattice with 
a phonon density of states $\rho(\omega)$, 

\begin{equation}
\rho ( \vec k,\vec k',+t ) = \frac{ \vec k.\vec k'}{2M}{\int_0^{\infty} d\omega \frac{\rho(\omega)}{\omega} [coth
(\omega/2kT) cos(\omega t) - i sin(\omega t)]}.
\end{equation}

In the short time approximation for $\tau$, $t$, and $t'$ the DDCS 
can be simplified using

\begin{equation}
\label{eqrho}
\rho ( k,k',+t ) \approx \rho ( k,k',0 ) + \vec k.\vec k'( \frac{-k T_{eff}}{2M}t^2 - \frac{i
t}{2M}),
\end{equation}
with
\begin{equation}
\label{eqlamb}
kT_{eff} = \frac{1}{2} \int_0^{\infty} d\omega \rho(\omega) \omega coth(\omega/2kT).
\end{equation}

Plugging \ref{eqrho} into \ref{eqclm}, we get the free gas formula \ref{eqfgm} with $T_{eff}$ instead of
$T$. Consequently, the crystal-lattice model lead to the free gas model, with an effective temperature, when
$\tau$, $t$, and $t'$ are small (short time approximation). This is the same effective temperature as Lamb \cite{lamb1939} derived for the Doppler broadening of a capture resonance in a solid in the weak binding limit.

\subsection{The uncoupled phonon approximation}

The UPA approximation proposed in \cite{nab2001} was an attempt to compute the resonant scattering kernel 
for a harmonic crystal. It neglected $t$ and $t'$ in the coupling terms
$\rho ( \vec k_i,\vec k_i,\tau-t'+t)$,
$\rho ( \vec k_i,\vec k_f,\tau-t' )$ and
$\rho ( \vec k_f,\vec k_i,\tau+t )$ in Equation \ref{eqclm} and applied a short time approximation to 
$\rho ( \vec k_i,\vec k_f,-t' )$ and $\rho ( \vec k_f,\vec k_i,+t )$
\footnote{A refinement of the model called MUPA (modified uncoupled phonon approximation)
computed  $\rho ( \vec k_i,\vec k_f,-t' )$ and $\rho ( \vec k_f,\vec k_i,+t )$
with a discrete phonon spectrum}. The correlation function
becomes 

\begin{eqnarray}
\chi^{UPA}_4(\tau,t,t') &  = & \exp \left( \rho ( \vec { \Delta k},\tau ) - \rho ( \vec { \Delta k},0 ) \right) \nonumber \\
  & \times  & \exp \left( \vec { k_i} \vec { k_f} [ \frac{-k T_{eff}}{2M}(t^2+t'^2) - \frac{i (t-t')}{2M}] \right).
\end{eqnarray}

The integral over $\tau$ and $t,t'$ can be performed separately. Since the first exponential is the Van-Hove function $\chi_2(\tau)$, the
differential cross section can be factored out as 

\begin{equation}
\frac{d^2\sigma}{d\Omega dE} = \frac{1}{4\pi}\frac{k_f}{k_i} 
\times S (\vec { \Delta k}, \Delta E) \times  \tilde{ \sigma } (\vec k_i, \vec k_f).
\end{equation}

The detailed mathematical expression for the so-called UPA cross section $\tilde{ \sigma } (\vec k_i, \vec k_f)$ can be found in \cite{nab2001}.
The UPA model separates the phonon dynamic and the nuclear interaction and provides a rather simple way to compute DDCS. 
However, within the UPA approximation, the 4-point correlation for small $\tau$, $t$, and $t'$ becomes 
 
\begin{equation}
\fl \chi^{UPA}_4(\tau,t,t') = 
\exp\{ \frac{-i}{2 M}[ \tau (\vec k_f - \vec k_i)^2 + ( t - t' ) \vec k_i \vec k_f] 
- \frac{kT_{eff}}{2 M} [\vec k_i \vec k_f ( t^2 + t'^2) + \tau^2 ( \vec k_f - \vec k_i )^2 ] \},
\end{equation}

which differs markedly from the required equation \ref{eqfgm} discussed in the previous section. 
Consequently, the DDCS within the UPA approximation does
not give the correct limit for small $t$ and $t'$.

\section{Proposed Model}

Our model seeks to get a factorization of the DDCS similar to the UPA model:

\begin{equation}
\label{eqfact}
\frac{d^2\sigma}{d\Omega dE} =  \frac{1}{4\pi} \frac{k_f}{k_i}  
\times S (\vec { \Delta k}, \Delta E ) \times \tilde{ \sigma }(\vec k_i, \vec k_f). 
\end{equation}

However, in order to get the correct limit at short time, we computed the $\tilde{ \sigma }$ term using the short time approximation of the correlation function (Equation \ref{eqfgm}), without the previous UPA approximation. 

For values of $\vec {\Delta k}  \neq \vec {0}$, the following notation is used,

\begin{equation}
\fl \Delta = \sqrt{\frac{4 k T}{M} \frac{k_i^2 k_f^2 - (\vec k_i \vec k_f)^2}{2 \vec{\Delta k}^2 }} \ \ x = 2 \ [ E_i - E_0 - (E_f - E_i) \ \frac{\vec k_i \vec \Delta k }{ \vec{\Delta k}^2 } - \frac{\vec k_i \vec
k_f}{2 M} ]\ /\ \Gamma.
\end{equation}

It is found that the differential cross section, including resonant, potential and resonant-potential interference 
can be calculated almost analytically and can be split into nuclear and transfer terms. 
The details of the demonstration are presented in the Appendix. The final result is:

\begin{equation}
\label{eqfinal}
\frac{d^2\sigma}{d\Omega dE} = \frac{1}{4\pi} \frac{k_f}{k_i} \times S (\vec { \Delta k}, \Delta E ) \times 
[\ \sigma_p + \sigma_m \frac{\Gamma_n}{\Gamma} \psi + \sqrt{ \sigma_p \sigma_m \frac{\Gamma_n}{\Gamma}} \chi\ ].
\end{equation}

For $\Delta \neq 0$ and $\vec {\Delta k}  \neq \vec {0}$,

\begin{equation}
\label{eqfinal2}
\psi = \frac{\sqrt{\pi}}{2} \times \frac{\Gamma}{\Delta} K(\frac{x\Gamma}{2\Delta},\frac{\Gamma}{2\Delta})
\ \ \ \chi = \sqrt{\pi} \times \frac{\Gamma}{\Delta} L(\frac{x\Gamma}{2\Delta},\frac{\Gamma}{2\Delta}).
\end{equation}

For $\Delta = 0$ and  $\vec {\Delta k}  \neq \vec {0}$

\begin{equation}
\psi = \frac{1}{1 + x^2} \ \ \ \chi = \frac{2 x}{1 + x^2}.
\end{equation}

where $K(x,y)$ and $L(x,y)$, are the real and imaginary part of complex complementary error function 
$w(z) = \exp[-(x+iy)^2]\ erfc\ [-i(x+iy)]$, related to the classic Voigt functions.

\begin{equation}
\label{eqfinal3}
K(x,y) = \Re[w(z)] = \frac{y}{\pi}\int_{-\infty}^{\infty}\frac{\exp (-t^2)}{(x-t)^2 + y^2} dt
\end{equation}

\begin{equation}
\label{eqfinal3}
L(x,y) = \Im[w(z)] = \frac{x}{\pi}\int_{-\infty}^{\infty}\frac{\exp (-t^2)}{(x-t)^2 + y^2} dt.
\end{equation}

When $\vec {\Delta k}  = \vec {0}$, the same calculations lead to Equation \ref{eqfinal} with :
\begin{equation}
\Delta = \sqrt{\frac{4 k T k_i^2}{2M}} \ \ x = 2 \ [ E_i - E_0 - \frac{ k_i^2 }{2 M} ]\ /\ \Gamma.
\end{equation}

We recognize the usual Doppler-broadened
cross section $\sigma^T$, so that the differential cross section becomes

\begin{equation}
\label{eqfact}
\frac{d^2\sigma}{d\Omega dE} =  \frac{1}{4\pi} \frac{k_f}{k_i}  
\times S (\vec { \Delta k}, \Delta E ) \times  \sigma^{T^*} (E^*), 
\end{equation}

where $T^{*}$ and $E^*$  depend on the cosine of the scattering angle $\mu = \cos \theta$:

\begin{equation}
\label{eqfinalresult}
T^*(\mu) = T \frac{k_f^2 ( 1-\mu^2 )}{\vec{\Delta k}^2 } \ \ \  E^* = E_i - (E_f - E_i) \ \frac{\vec k_i \vec \Delta k }{ \vec{\Delta k}^2 } - \frac{\vec k_i \vec
k_f}{2 M},
\end{equation}

and for $\vec {\Delta k}  = \vec {0}$

\begin{equation}
T^* = T  \ \ \  E^* = E_i  - \frac{\vec k_i^2}{2 M}.
\end{equation}

Equation \ref{eqfinalresult} is valid only when $E_i - (E_f - E_i) \ \frac{\vec k_i \vec \Delta k }{ \vec{\Delta k}^2 } - \frac{\vec k_i \vec k_f}{2 M}$ is positive. The previous equation can be further simplified noting that $T^* \approx T [(1+\mu) /2][1+\Delta E/(2 E_i)] \approx T (1+\mu) /2 $ and when $1-\mu >> |\vec { \Delta k}|/|\vec k_i|$ we have $\frac{\vec k_i \vec \Delta k }{ \vec{\Delta k}^2 } \approx -1/2$ and this approximation produces satisfactory results when used over the whole range of values of $\mu$. Making these approximations, we get a simple formula

\begin{equation}
\label{eqfinalresult2}
\frac{d^2\sigma}{d\Omega dE} =  \frac{1}{4\pi} \frac{k_f}{k_i}  
\times S (\vec { \Delta k}, \Delta E ) \times  \sigma^{T(1+\mu)/2} \left( [E_i+E_f]/2 -E_i\mu/A \right). 
\end{equation}

Equation \ref{eqfinalresult2} provides a simple way to compute the differential cross section
within the free gas model. It also gives an approximate way to account for solid state effects
by using the known scattering function $S (\vec { \Delta k}, \Delta E )$ for the harmonic crystal.
Note that  Equation \ref{eqfinalresult2} was demonstrated for a single resonance and it is not known if
this formula can be generalized to any form of free scattering amplitude.

\section{Application to neutron resonant scattering in UO$_2$}

Low-enriched uranium oxide UO$_2$ is widely used as nuclear-reactor fuel.
With the present model, the scattering kernel
of $^{238}$U in UO$_2$  has been calculated near the first resonance at 6.67 eV 
by numerical integration of the differential cross section over scattering angles. 
The weighted phonon spectrum for $^{238}$U in UO$_2$ has been taken from the measurement
of Dolling et al. \cite{dolling1965}. The spectrum features two acoustic modes around 14 and 21 meV.

The $S(\vec { \Delta k}, \Delta E)$ Van-Hove scattering function has been computed using the usual 
phonon expansion methods. The $^{238}$U resonance parameters evaluated by Moxon and Sowerby \cite{moxon1994} were used.

The present model has been compared with the classic free gas kernel
published by Blackshaw and Murray \cite{blackshaw1967} and studied by
Ouisloumen and Sanchez \cite{ouisloumen1991}. Fig. 1 and 2 show the resulting scattering kernel
at 300 K at two incident neutron energies, 6.674 eV (the peak of the first resonance)
and 6.520 eV which is the minimum of the cross section due the interference
between the resonant and potential amplitudes. Compared to the UPA model (see Figures in Reference \cite{nab2001}),
the present crystal model gives results closer to the free gas model. The only distinctive features are the
small peaks around the elastic peak, associated with the one-phonon excitation of the acoustic modes. 
Note that when the $S(\vec{\Delta k}, \Delta E)$ of the FGM is used in Equation \ref{eqfinal}, the
numerical computations give the same results as Ouisloumen and Sanchez.

\begin{center}
\begin{figure}[h] 
\center
\includegraphics{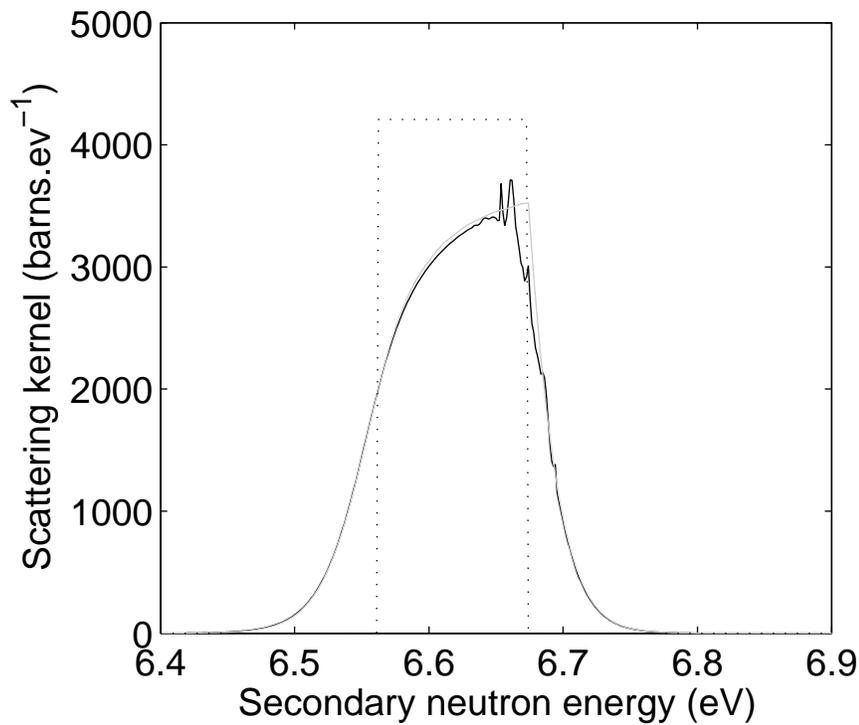}
  \caption{\label{comparison1} Comparison of scattering kernels calculated with the present formalism (black line) and the free gas model of Ouisloumen and Sanchez \cite{ouisloumen1991} (grey line) at incident neutron energy of 6.674 eV. The dotted line is the kernel with a static model.} 
\end{figure}
\end{center}

\begin{center}
\begin{figure}[h]
\center 
\includegraphics{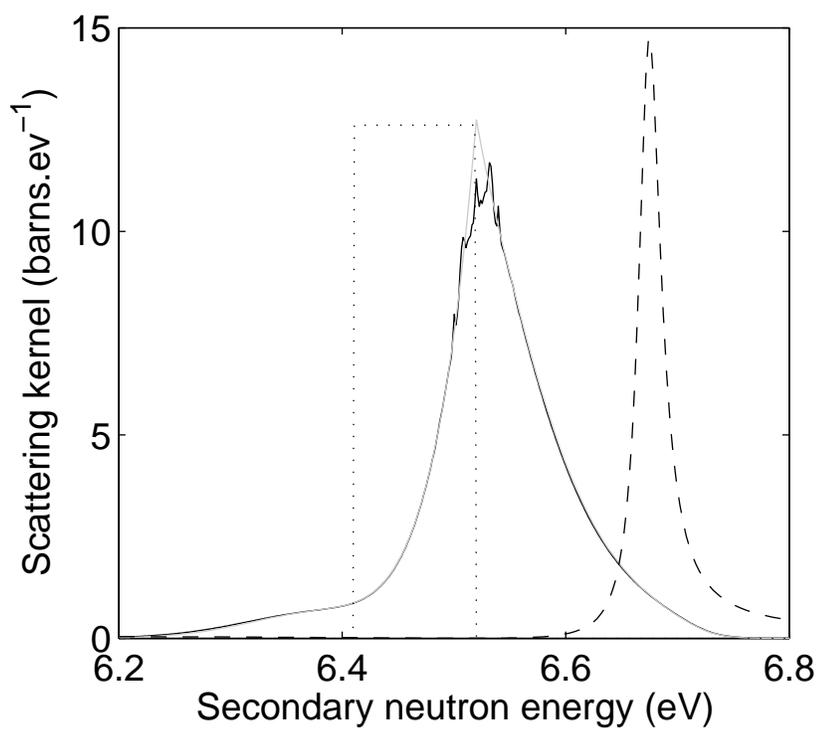} 
  \caption{\label{comparison2} Comparison of scattering kernels calculated with the present formalism (black line) and the free gas model of Ouisloumen and Sanchez \cite{ouisloumen1991} (grey line) at incident neutron energy of 6.520 eV. The dotted line is the kernel with a static model and the dashed line is the scattering cross section ($\sigma_s / 100$ in barns). } 
\end{figure}
\end{center}

\clearpage

\section{Conclusion}

An approximate formula is proposed to treat solid state effects in neutron-crystal interactions.  The DDCS is the product of 
a Doppler-broadened scattering cross section and the usual Van-Hove scattering function.

\begin{equation}
\label{eqconcl1}
\frac{d^2\sigma}{d\Omega dE} =  \frac{1}{4\pi} \frac{k_f}{k_i}  
\times S (\vec { \Delta k}, \Delta E ) \times  \sigma^{T(1+\mu)/2} \left( [E_i+E_f]/2 -E_i\mu/A \right). 
\end{equation}
This formula gives the correct free-gas limit for short-time range (high-temperature cases).
A rigorous calculation of the DDCS within the crystal lattice model is still an open issue. 
The present work develops the case of a single isolated resonance but further generalizations
to multilevel forms of collision matrix may be possible. 

 In reactor applications, the model predicts small solid-state effects in the scattering of neutrons in UO$_2$ at 300K, contrary to previous studies using the UPA approximation. Measurements of the secondary spectrum of neutrons scattered elastically in resonances would be valuable to check existing scattering models.

\ack
It is a pleasure to acknowledge fruitful discussions with R. Dagan, C. R. Lubitz,
and A. Santamarina.

\appendix

\section{Derivation of Equation \ref{eqfinal} }

For $\vec {\Delta k}  \neq \vec {0}$, $\chi_4$ in Equation \ref{eqfgm} can be transformed into

\begin{eqnarray}
\chi_4  & = & 
\exp\{ \frac{-i}{2 M}[ \left(\tau - \frac{\vec k_i \vec \Delta k (t-t')}{\vec{\Delta k}^2} \right) 
\vec \Delta k ^2 + (t-t') \vec k_i \vec k_f ] \nonumber \\
& - & \frac{kT}{2 M} [ \left(\tau - \frac{\vec k_i \vec \Delta k (t-t')}{\vec{\Delta k}^2} \right)  ^2 \vec \Delta k^2 - 
\frac{k_i^2 k_f^2 - (\vec k_i \vec k_f)^2}{\vec{\Delta k}^2} (t-t')^2 ] \}.
\end{eqnarray}

Using the variable $u = \tau - \vec k_i \vec \Delta k (t-t') / \vec{\Delta k}^2$, 
we recognize $\chi_2$  the pair correlation function for a free gas: 
\begin{equation}
 \chi_2(u) = \exp\{ \frac{-i}{2 M} u \vec{\Delta k}^2 
- \frac{kT}{2 M} u^2 \vec{\Delta k}^2  \}. 
\end{equation}

Therefore, the Fourier transform of $\chi_4$ (integration over $\tau$) will lead to the
$S(\vec \Delta k, \Delta E)$ terms with a phase factor to account for the change of variable 
$\tau \rightarrow u$. The DDCS takes the form of Equation \ref{eqfact} with
 
 \begin{eqnarray}
 \label{eqint2}
 \tilde{ \sigma } & = & \sigma_m \frac{\Gamma_n}{\Gamma} \frac{\Gamma^2}{4}  
 \int_{0}^{\infty}\int_{0}^{\infty} e^{ -z t } e^{-z^* t'}\ \phi(t-t') dt dt',  
\end{eqnarray}

with $z = -i \left( E_i-E_0 - \left( E_f-E_i \right ) \frac{\vec k_i \vec \Delta k}{\vec{\Delta k}^2 } - \frac{\vec k_i \vec k_f}{2 M} \right) +\frac{ \Gamma}{2} $ and 

\begin{equation}
\phi(t) = \exp \left( - \frac{kT}{2 M} \frac{k_i^2 k_f^2 - (\vec k_i \vec k_f)^2}{\vec{\Delta k}^2 } t^2  \right).
\end{equation}

To compute the double Laplace transform, the following identity is used:

\begin{equation}
\int_0^{\infty} \int_0^{\infty} e^{-zt - z^* t'} \phi(t-t') dt dt' =   
\frac{1}{z + z^*}\int_0^{\infty} e^{-zt} \phi(t) dt + \int_0^{\infty} e^{-z^* t} \phi^*(t) dt.
\end{equation}

Equation \ref{eqint2} becomes 

\begin{eqnarray}
\label{eqint3}
\tilde{ \sigma } & = & \sigma_m \frac{\Gamma_n}{4}  \int_0^{\infty} 
 e^{ - i \left( E_i-E_0 - \left( E_f-E_i \right ) \frac{\vec k_i \vec \Delta k}{\vec{\Delta k}^2 } - \frac{\vec k_i \vec k_f}{2 M} \right)t +\frac{ \Gamma t}{2}} e^{\left( - \frac{kT}{2 M} \frac{k_i^2 k_f^2 - (\vec k_i \vec k_f)^2}{\vec{\Delta k}^2 } t^2  \right)} dt  \nonumber \\
& + &  c.c. 
\end{eqnarray}

c.c denotes the conjugate complex. This equation is then reduced into a single integral which can be further simplified using

\begin{equation}
\label{eqc}
\int_{0}^{\infty} e^{-iat -bt^2} dt = 
\sqrt{\frac{\pi}{b}} \exp  \frac{-a^2}{4b}  erfc\ [\frac{i a}{\sqrt{4b}} ]
\end{equation}

We recognize in \ref{eqint3} and \ref{eqc} the real of the complex complementary error function. Equations \ref{eqfinal}, \ref{eqfinal2} and \ref{eqfinal3} are obtained in a similar way when the potential and potential-resonant interference terms are accounted for.


\section*{References}
\begin{thebibliography}{00} 

\bibitem{wigner1944}
{Wigner E P and Wilkins E J}
1944 {AECD-2275, Clinton Laboratory}

\bibitem{blackshaw1967}
{Blackshaw G L and Murray R L}
1967 {\it Nucl. Sci. Eng. } {\bf 27}, 520--532

\bibitem{ouisloumen1991}
{Ouisloumen M and Sanchez R}
1991 {\it Nucl. Sci. Eng. }{\bf 107}, 189--200




\bibitem{roth4}
{Rothenstein W}
2004 {\it Ann. Nucl. Energy } {\bf 31}, 1, 9--23

\bibitem{vanhove1954}
{Van Hove L}
1954 {\it Phys. Rev. }{\bf 95}, 249

\bibitem{trammel1962}
{Trammel G T}
1962 {\it Phys. Rev.} {\bf 126}, 1045

\bibitem{word1981}
{Word R E and Trammell G T}
1981 {\it Phys. Rev. } {\bf B24}, 2430

\bibitem{kazarnovski1962}
{Kazarnovskii M V and Stepanov A V}
1962 {\it Sov. Phys. JETP, } {\bf 15}, 2, 343--349

\bibitem{nab2001}
{Naberejnev D G}
2001 {\it Ann. Nucl. Energy} {\bf 28}, 1

\bibitem{dolling1965}
{Dolling G et al.}
1965 {\it  Can. Journ. Phys.} {\bf 43}

\bibitem{moxon1994}
{Moxon M and Sowerby M} 
1994 {\it Summary of the Work of the NEANDC Task Force on U-238,} Nuclear Energy Agency, OECD report

\bibitem{lamb1939}
{Lamb W E} 
1939 {\it Phys. Rev.} {\bf 55}, 190

\bibitem{shamaoun1990}
{Shamaoun A I and Summerfield G C}
1990 {\it  Ann. Nucl. Energy} {\bf 17}, 229

\end{thebibliography}
\end{document}